\documentclass[]{spie}  
\usepackage[]{graphicx}

\usepackage{color}
\usepackage{bm}

\usepackage[english]{babel}
\usepackage{amsmath}
\usepackage{amssymb}
\usepackage{units}
\usepackage{upgreek}
\usepackage[colorlinks=true, allcolors=blue]{hyperref}
\usepackage{ctable}
\usepackage{footnote}
\usepackage{bbold}
\usepackage{graphicx}
\usepackage{subcaption}
\usepackage{textcomp}
\usepackage[absolute,overlay]{textpos}
\usepackage{helvet}
\usepackage{svg}
\newcommand{\orcid}[1]{\href{https://orcid.org/#1}{\includesvg[width=11pt]{orcid}}}

\title{Selective laser etching of displays: Closing the gap between optical simulations and fabrication} 

\author{Martin Wimmer,\supit{a} Myriam Kaiser,\supit{a} Jonas Kleiner,\supit{a}\;\orcid{0000-0003-1985-0382}, Jannis Wolff,\supit{a} Max Kahmann,\supit{a} and Daniel Flamm\supit{a}\;\orcid{0000-0003-2438-5980}
\skiplinehalf
\supit{a}TRUMPF Laser- und Systemtechnik GmbH, Johann-Maus-Str.\,2, 71254 Ditzingen, Germany
}

\authorinfo{Further author information:\\ E-Mail: \href{mailto:myriam.kaiser@trumpf.com}{martin.wimmer@trumpf.com}.}

\begin{document} 
  \maketitle 

\begin{abstract}
Simulations and measurements on selective laser etching of display glasses are reported. By means of a holographic 3D beam splitter, ultrashort laser pulses are focused inside the volume of a glass sample creating type III modifications along a specific trajectory like pearls on a string. Superimposed by a feed of the glass sample a full 3D area of modifications is achieved building the cornerstone for subsequent etch processes. Based on KOH the modifications are selectively etched at a much higher rate compared to unmodified regions resulting in a separation of the glass along the trajectory of modifications. For gaining further insight into the etch process, we perform simulations on this wet chemical process and compare it to our experimental results.
\end{abstract}



\section{INTRODUCTION}
\label{sec:intro}  
\begin{textblock*}{16cm}(2.67cm,1cm) 
   \centering
  \tiny \textsf{Martin Wimmer, Myriam Kaiser, Jonas Kleiner, Jannis Wolff, Max Kahmann, Daniel Flamm, "Selective laser etching of displays: closing the gap between optical simulations and fabrication," Proc. SPIE 12872, Laser Applications in Microelectronic and Optoelectronic Manufacturing (LAMOM) XXIX, 1287206 (12 March 2024); \url{https://doi.org/10.1117/12.3000501}.}
\end{textblock*}

\begin{textblock*}{17cm}(2.25cm,25.25cm) 
   \centering \small 
   \textsf{
   © 2024 Society of Photo‑Optical Instrumentation Engineers (SPIE). One print or electronic copy may be made for personal use only. Systematic reproduction and distribution, duplication of any material in this publication for a fee or for commercial purposes, and modification of the contents of the publication are prohibited. \\
   Laser Applications in Microelectronic and Optoelectronic Manufacturing (LAMOM) XXIX © 2024 SPIE.  \url{https://doi.org/10.1117/12.3000501}.}
\end{textblock*}
The displays of mobile devices must withstand extreme conditions not only due to the large variety of environmental conditions on earth, but they also have to bear up against shock impacts challenging the structural mechanics of the cover glass. Consequently, the robustness of the transparent substrate is pushed by means of material science towards new limits, but also the form of the edge is of importance as it influences the impact and bending strength of the glasses.\cite{marjanovic2019edge, bukieda2020study, flamm2021protecting} Besides finding theoretically the perfect edge shape, the producibility is likewise mandatory. Laser-based glass cutting concepts \cite{ahmed2008display, tsai2014internal, kumkar2016ultrafast}---as with mechanical scribe and break processes\cite{zhou2006tool,nisar2013laser}---primarily leave a straight edge, representing the main weak point of the substrate.\cite{flamm2021protecting}
\par 
Recently, we introduced an approach based on photonic shaping tools.\cite{flamm2021protecting, flamm2023photonic} This combines the advantages of laser-based cutting of glass with concepts for shaping the edges. Here, a holographic 3D beam splitter\cite{valle2012analytic,zhu2014three,flamm2021structured} allows to simultaneously distribute a large number of foci inside the bulk of transparent media. Using adapted parameters from ultrashort pulsed lasers result in type-III-regime modifications\cite{glezer1997ultrafast,itoh2006ultrafast} which, when placed densely inside the volume, define a boundary between two regions of a glass sample. Finally, both parts can be separated mechanically, thermally, e.g., by a CO$_2$ laser\cite{kaiser2023tailored} or by selectively etching modifications\cite{bellouard2004fabrication,hermans2014selective,gottmann2017selective,kaiser2019selective}. With the latter approach we take advantage from the fact that laser-induced type-III-regime modifications exhibit a higher etchability than the ground material. Selective laser etching (SLE) is particularly profitable if economies of scale need to be achieved, i.e., if a large number of items have to be produced at the same time. In these cases, it is reasonable to use aqueous solutions of chemicals such as potassium hydroxide (KOH),\cite{hermans2014selective} sodium hydroxide (NaOH)\cite{casamenti2021few} or even hydrofluoric acid (HF).\cite{bellouard2004fabrication}
\par
In what follows, we introduce an approach for simulating the SLE process with the aim to predict the substrate's edge shape and to optimize the required focus distribution. This is where the most prominent advantage of the photonic shaping tool comes into play, which is the ability to freely distribute spots in the working volume of a focusing unit.\cite{flamm2023photonic} With this design-degree-of-freedom we control the shape of the entirety of type-III-regime modifications at which the etch solution is penetrating the substrate. The numerics presented enable the prediction of edge shapes and process times for processing with different etch rates. Additionally, we present the impact of assist etch lines facilitating etching and solving the detachment problem for separating complex inner contours.\cite{kleiner2023laser} Finally, we extend the SLE simulations to three spatial dimensions allowing to optimize focus shapes when complex contours are to be processed such as sub-millimeter scaled vias.

\section{Simulation of the selective laser etching}\label{sec:sims}
\begin{figure}
    \centering
    \hfill
    \begin{subfigure}{6cm}
        \includegraphics[width=\textwidth]{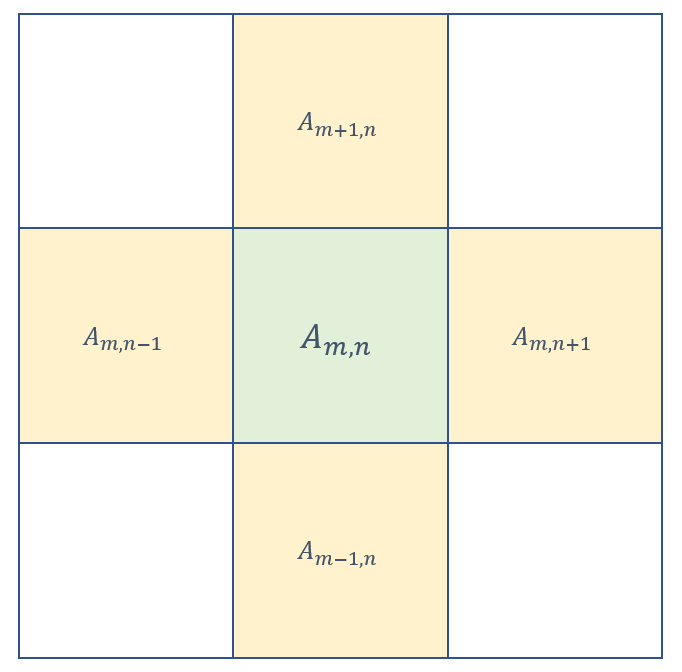}
    \end{subfigure}
    \hfill
    \begin{subfigure}{3.5cm}
        \includegraphics[width=\textwidth]{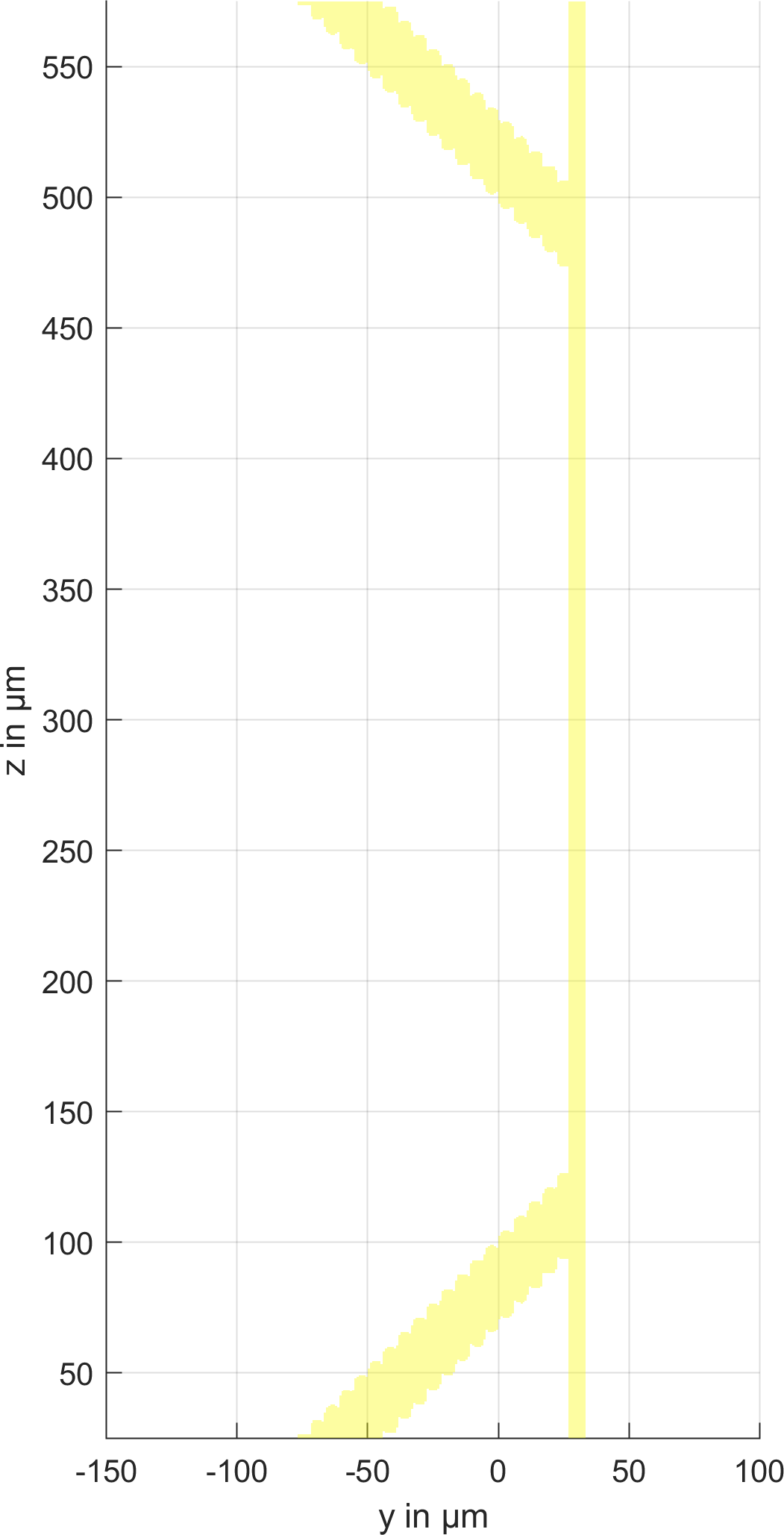}
    \end{subfigure}
    \hfill \vspace{1mm}
    \caption{Sketch of the mesh model in the simulation (left panel). Only nearest neighbors are included. The full simulation domain in 2D is approximately 250 $\upmu$m $\times$ $625$ $\upmu$m large, for simulating the chamfer of a 500 $\upmu$m thick glass sample. The entirety of modifications generated by a single laser pulse propagating parallel to the $z$-axis is highlighted in yellow. Here, the focus shape generates modifications along a $45^\circ$ chamfer trajectory and, additionally two assist lines facilitating the edge process. For processing and contour cutting the focus distribution is moved relatively to the workpiece in $x$-direction.\cite{flamm2021protecting} }
    \label{fig:EtchingModel}        
\end{figure}

Our mathematical model for simulating the etching process is based on a cellular automaton\cite{weimar1994class} for solving the problem of a reaction diffusion equation in 2D and also 3D. In general, such models can cover not only isotropic etching but also be extended to anisotropic \cite{zhu2000simulation}. In what follows, we assume a full isotropic behavior, which agrees very well with experimental results. The initial condition of the numerical model is a homogeneous glass sample, with a constant glass concentration $A_{m,n}$ for each lattice site $m$ along the horizontal and $n$ along the vertical axis. In the 3D case, we add another spatial lattice site index $o$. Here, $A_{m,n}=1$ describes a point in the simulation domain, where the glass concentration is 100\,\% and the opposite $A_{m,n}=0$ a point, at which the glass is completely etched (see Fig.~\ref{fig:EtchingModel}). For the simulations, we differ only between these two binary states $A = 0$ and $A = 1$ but for visualization of the dynamics the continuous concentration is shown. \par
The etching process itself is driven by surrounding areas, where each next neighbor (NN) contributes to the etching at site $m,n$ with an etching rate $B$, which is adapted to experimentally determined etch rates of bare glass samples. Including the etch selectivity $S_{m,n}$ at each lattice site $m$, $n$ completes the evolution equation
\begin{equation}
    A^{t + \Delta t}_{m,n} = A^{t}_{m,n} - B \operatornamewithlimits{\sum}_{m\prime,n\prime \in \mathrm{NN}} (1 - A^{t}_{m\prime,n\prime}) - S_{m,n} \cdot \left(1 - A^{t}_{m,n}\right).
    \label{eq:MeshEq}
\end{equation}
In the same way as Eq.~\eqref{eq:MeshEq}, we can also expand the equation to three spatial dimensions
\begin{equation}
    A^{t + \Delta t}_{m,n,o} = A^{t}_{m,n,o} - B \operatornamewithlimits{\sum}_{m\prime,n\prime,o\prime \in \mathrm{NN}} (1 - A^{t}_{m\prime,n\prime,o\prime}) - S_{m,n,o} \cdot \left(1 - A^{t}_{m,n,o}\right),
    \label{eq:MeshEq3D}
\end{equation}
which will be later used for three dimensional simulations.\par
For providing a conclusive picture, we want to present a continuous counterpart of the 2D evolution equation above 
\begin{equation}
    \frac{1}{B}\frac{\partial}{\partial t} A(t,x,y) = \left( \frac{\partial^2}{\partial x^2} + \frac{\partial^2}{\partial y^2} \right) A(t,x,y) - \left( 4 + \frac{S(x,y)}{B} \right) \left(1 - A(t,x,y) \right).
\end{equation}
Here, the first term on the right hand side stands for the diffusive part, the second term resembles the reaction driven by the surrounding reagent concentration $1 - A(t,x,y)$ and the local etch selectivity $S(x,y)$ raised by the laser inscribed modifications. The etch rate $B$ of the homogeneous glass is here only a trivial scaling of the temporal evolution. While the continuous counterpart provides insight into the physical model behind the etching process, in the following we will focus on the discrete model for determining numerically the temporal evolution.\par
For tailoring the etching behavior to a specific form of the chamfer, in the experiment glass is modified by focusing ultra short laser pulses at a position $(x,y)$. The high intensity of the laser pulse triggers a cascade of nonlinear processes resulting in a type-III-regime modification of the glass\cite{glezer1997ultrafast,itoh2006ultrafast}. Here, the modification volume is determined by the size of the focal spot and consequently, we assume in experiments a spot diameter of $2.5\,\upmu\mathrm{m}$ in transversal direction and the resulting Rayleigh length $z_\mathrm{R} \approx 13\,\upmu\mathrm{m}$ for an approximately $M^2=1$ beam in longitudinal direction at $\lambda = 1030\,\mathrm{nm}$. Furthermore, the focusing of the ultrashort pulses leads to extreme mechanical stress inside the glass resulting ultimately in micro explosions. In our heuristic model, we assume a much higher etch selectivity inside the volume of the focus\cite{flamm2021protecting,kaiser2022chamfered,kaiser2019selective}. We model these modifications by Super-Gaussian distributions for 2D modifications
\begin{equation}
    S_{m,n}=S_0 \cdot \exp\left(-\frac{\left|n\right|^N}{\left|w_\mathrm{n}\right|^N}-\frac{\left|m\right|^N}{\left|w_\mathrm{m}\right|^N}\right),
    \label{eq:Defects}
\end{equation}
which is analogously expanded to 3D.\par
Compared to binary model of modifications, the Super-Gaussian distribution in Eq.~\eqref{eq:Defects} features the advantage, that the modification is a smooth distribution on the grid and in this way, aliasing artifacts as in case of single lattice site modifications are prevented. The Super-Gaussian exponent $N$ controls the sharpness of the defects and typically, we use an exponent of $N = 4$ for our simulations.\par
To further refine the model, the shape and extent of the individual voids could be deduced from measurements of type-III-regime modifications caused by single Gaussian spots, see Refs.~\citenum{grossmann2016transverse, bergner2018spatio}. Additionally, for determining the degree of etch selectivity $S_0$, we refer to, for example, Hermanns \textit{et al.} where etch rates were measured from polarization contrast microscopy \cite{hermans2014selective}.
\par
As an initial condition of the simulation we reduce the glass concentration of the upmost $m = 0$ and bottommost $m = M$ row to $A_{n,0} = A_{n,M} = 0$, which corresponds to the interface between the etching solution and the (not necessarily) plane glass surface.\par
During the etching simulation, glass is homogeneously removed until the solution gets in contact with a modified region. In this case, the process drastically speeds up and the time before contact with the next modified areas is extremely reduced. In this way, all modifications are etched before a significant amount of glass is removed.

\section{Comparison of experimental and numerical results}
\begin{figure}
    \centering
    \hfill
    \begin{subfigure}{4cm}
        \includegraphics[width=\textwidth]{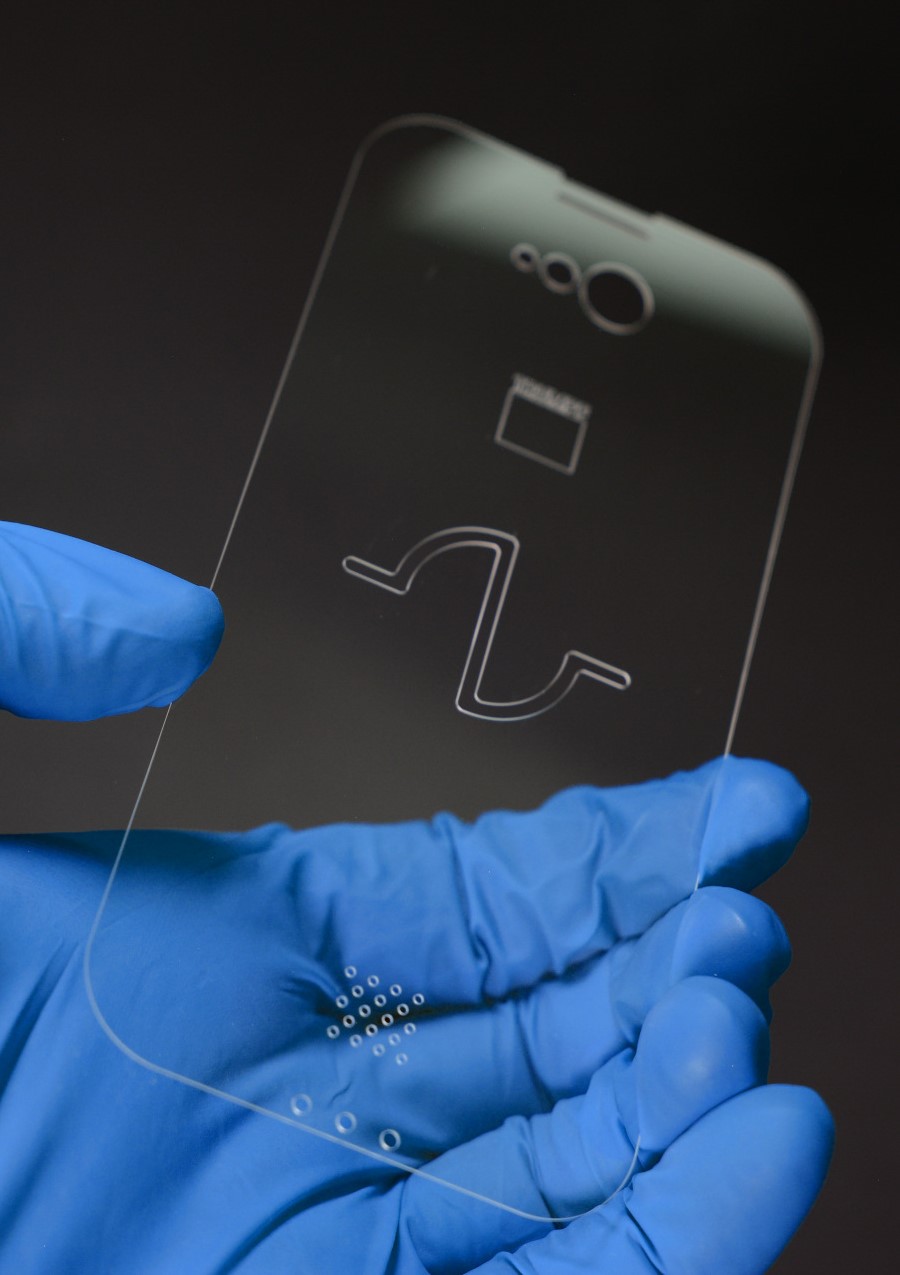}
    \end{subfigure}
    \hfill
    \begin{subfigure}{6.5cm}
        \includegraphics[width=\textwidth]{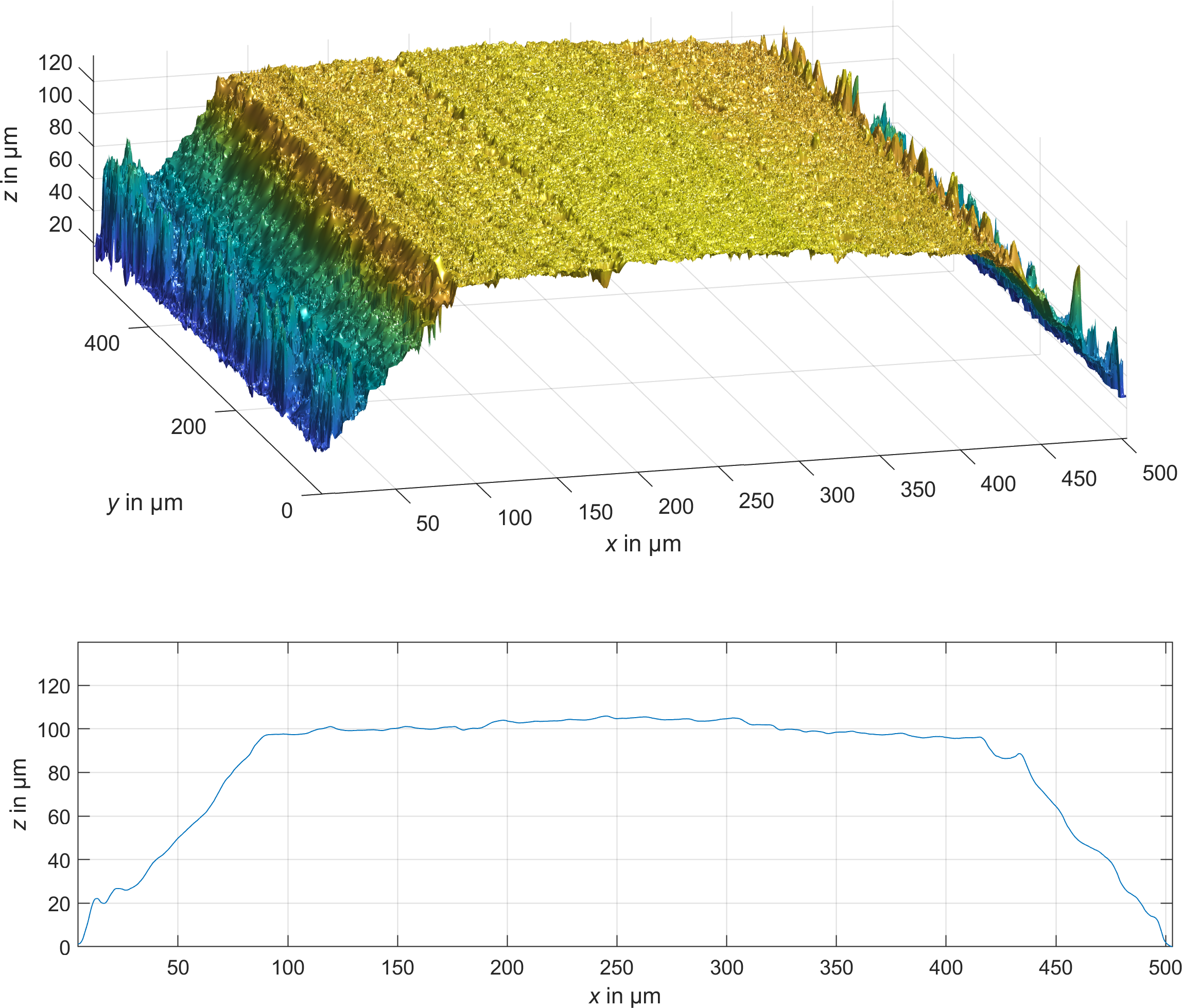}
    \end{subfigure}
    \hfill
    \begin{subfigure}{4.5cm}
        \includegraphics[width=\textwidth]{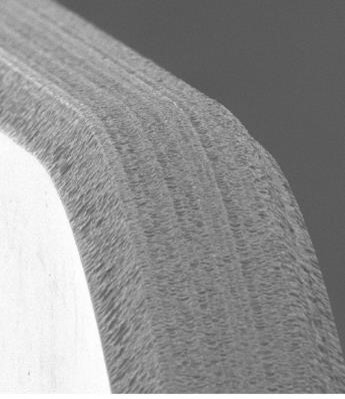}
    \end{subfigure}
    \hfill \vspace{1mm}
    \caption{Photography of the selectively laser etched sample (left) with laser scanning microscope measurements (center) of the chamfer and averaged cross section. The right panel depicts a scanning electron microscope image of the etched chamfer.}
    \label{fig:FotoShowSample}
\end{figure}

\begin{figure}
    \centering
    \hfill
    \begin{subfigure}{3cm}
        \includegraphics[width=\textwidth]{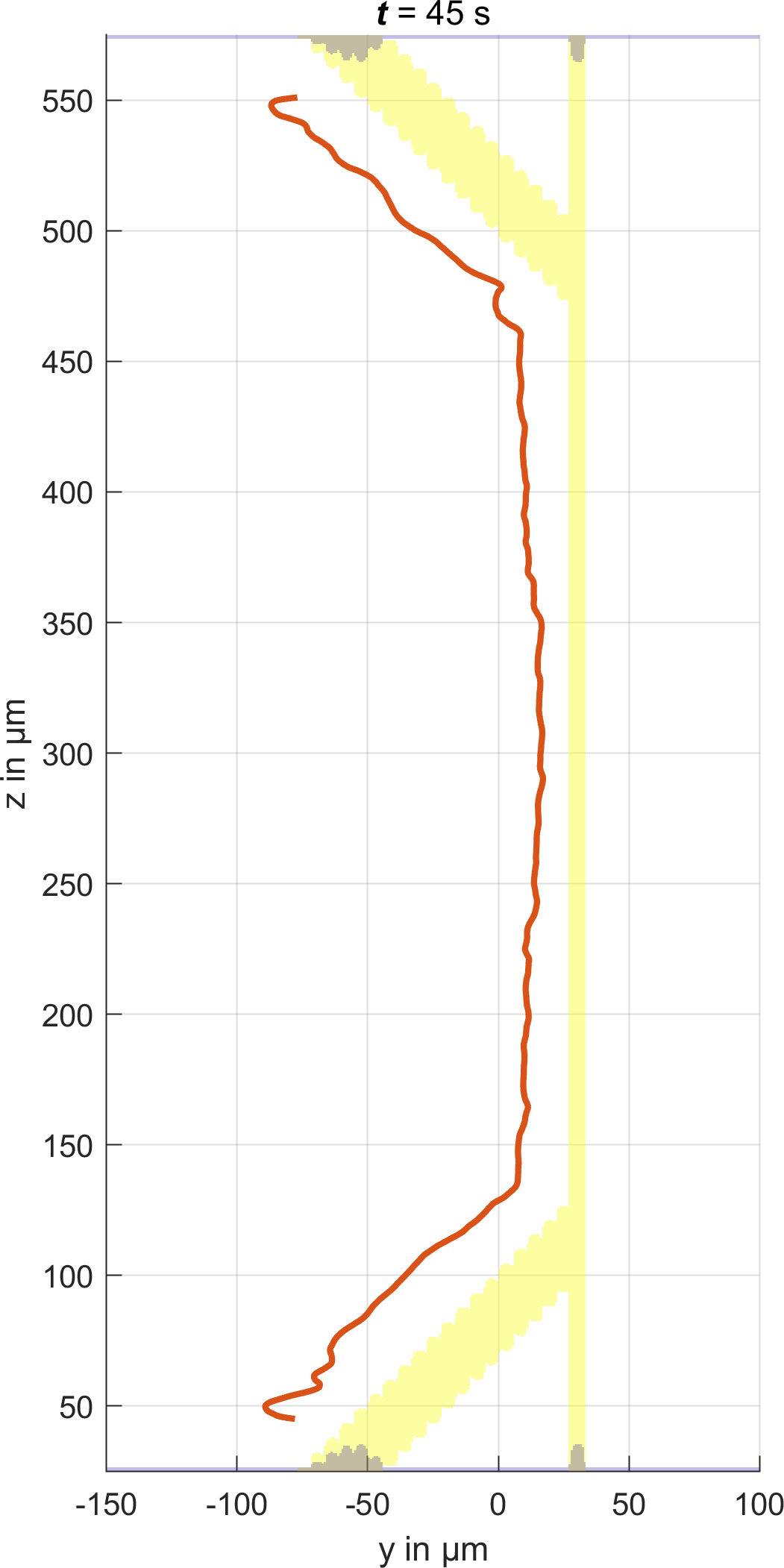}
    \end{subfigure}
    \hfill
    \begin{subfigure}{3cm}
        \includegraphics[width=\textwidth]{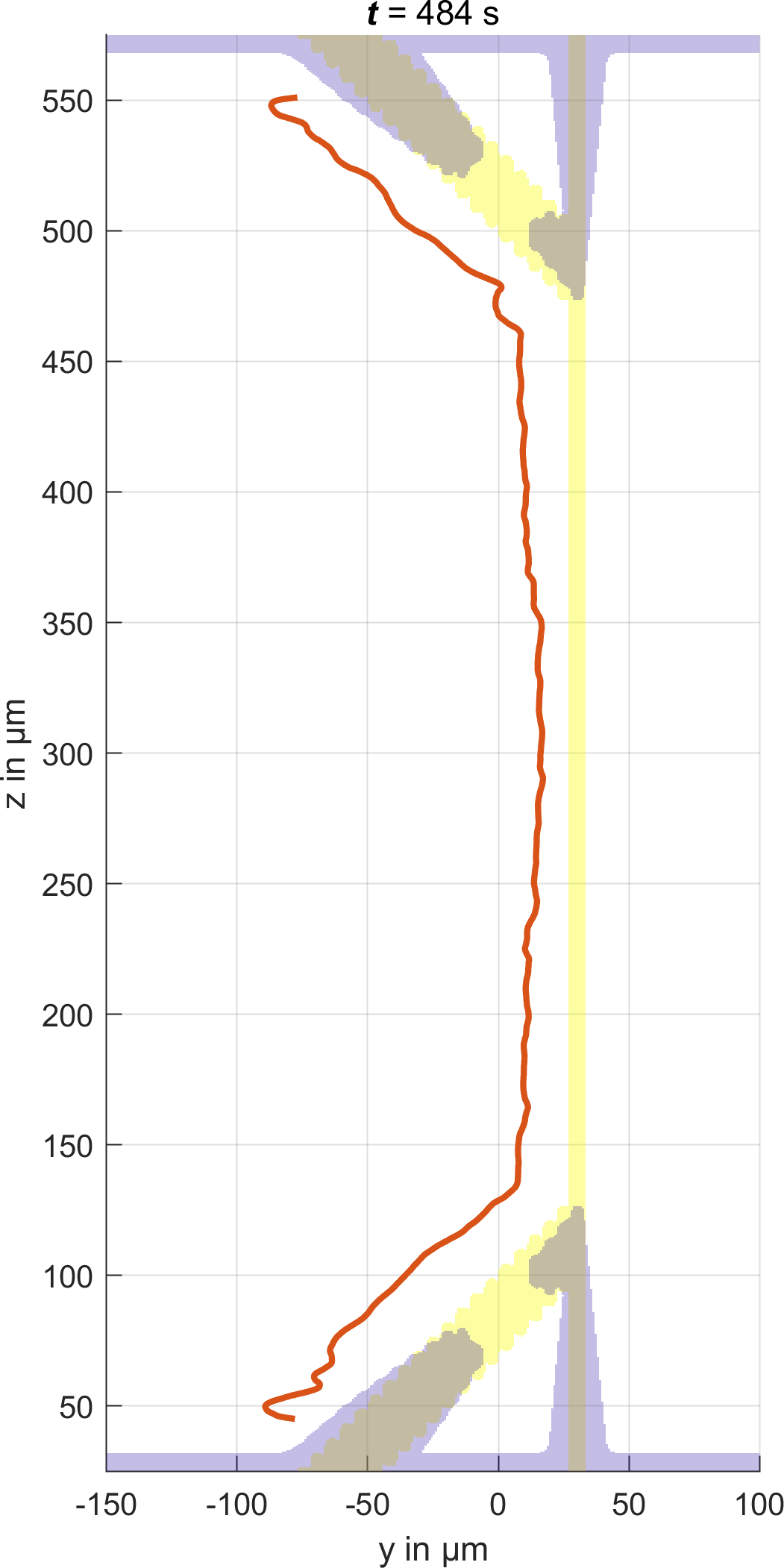}
    \end{subfigure}
    \hfill
    \begin{subfigure}{3cm}
        \includegraphics[width=\textwidth]{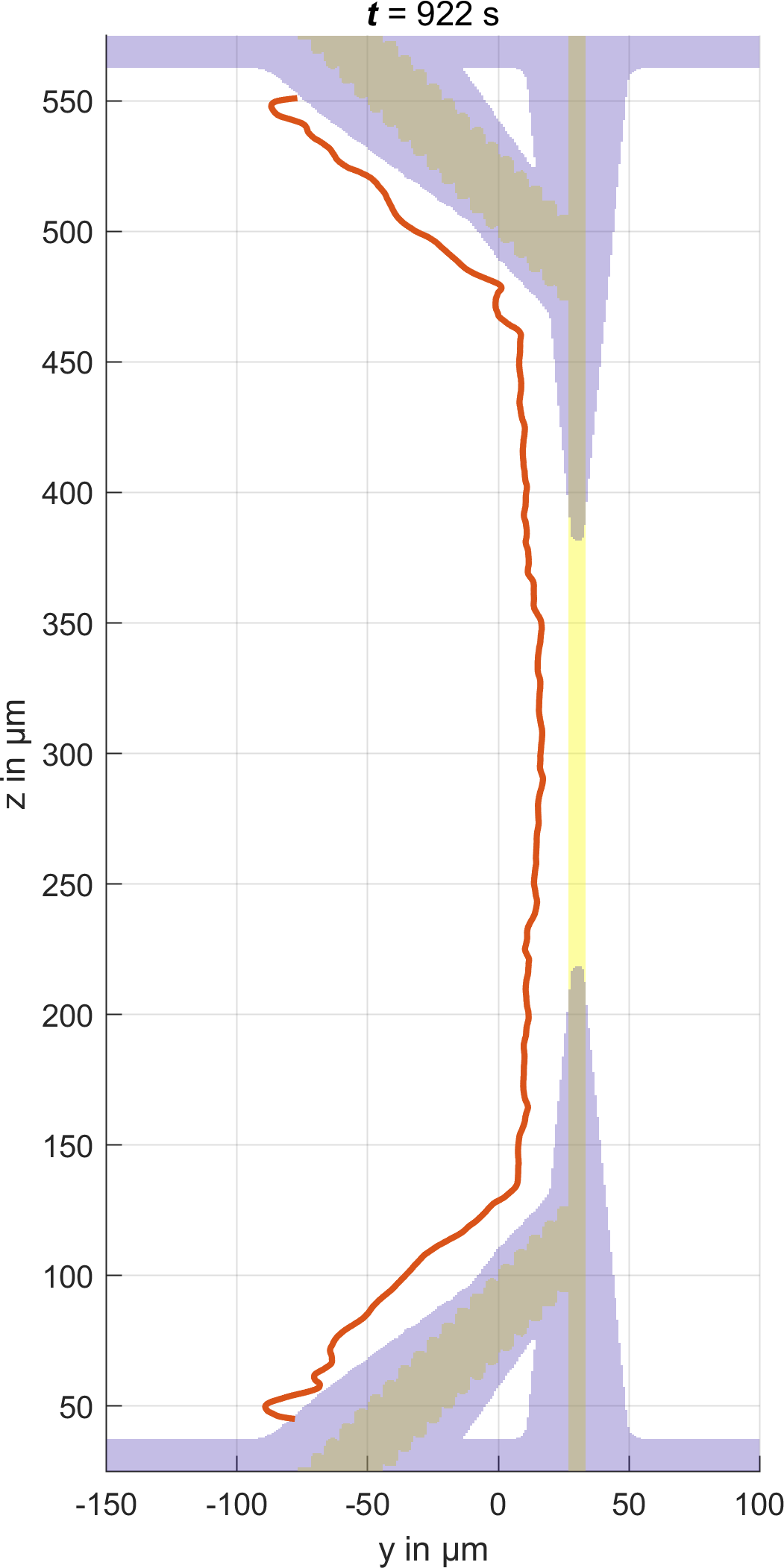}
    \end{subfigure}
    \hfill
    \begin{subfigure}{3cm}
        \includegraphics[width=\textwidth]{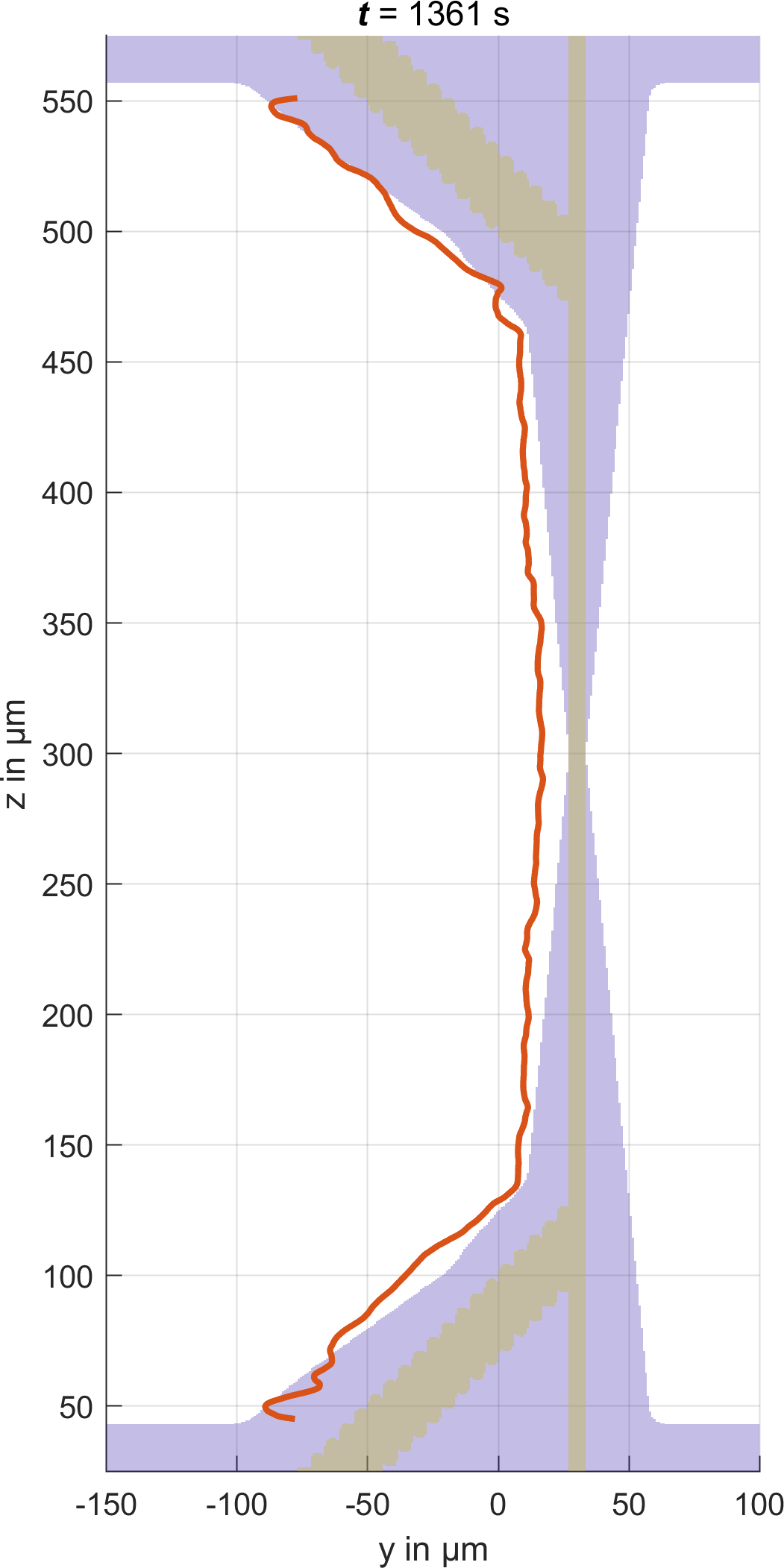}
    \end{subfigure}
    \hfill
    \begin{subfigure}{3cm}
        \includegraphics[width=\textwidth]{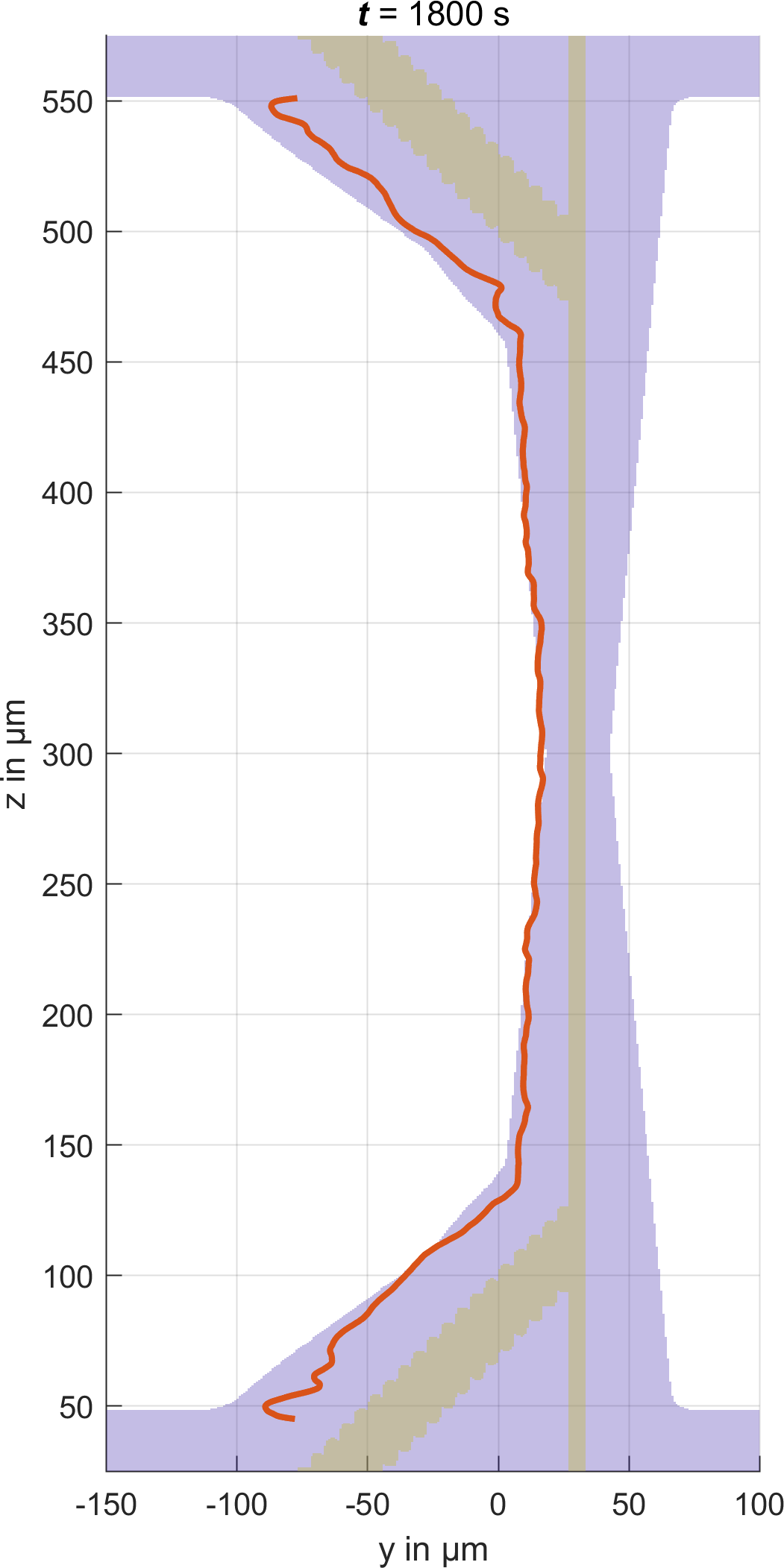}
    \end{subfigure}
    \hfill \vspace{1mm}
    \caption{Simulation of the selective laser etching for five different time steps of the process until the sample is etched down to a total thickness of 500$\,\upmu$m. The design includes two 45$\,^\circ$ chamfers at the top and bottom side of a 500$\,\upmu$ thick glass sample. For improving the etch results, two additional vertical channels are modified, which lead to a fast transport of the etching solution to the central part of the sample. The shape of the simulated chamfer fits well to the experimental results (solid red curve) based on the LSM measurements shown in Fig.~\ref{fig:FotoShowSample}. The yellow shaded region indicates the modified region, where spots according to Eq.~\eqref{eq:Defects} are placed.}
    \label{fig:SimulationResults2D}
\end{figure}

In the experiments, we use 3$\,$ps pulses from a \href{https://www.trumpf.com/en_INT/products/laser/short-and-ultrashort-pulse-laser/trumicro-series-2000/}{TruMicro Series 2000 laser} in burst-mode with a total pulse energy of $<$ 300$\,$µJ. The pulses of the laser pass a liquid-crystal-on-silicon-based spatial light modulator (SLM), which splits the light according to our principle of the holographic 3D beam splitter in several directions, each focusing inside a Corning\textsuperscript{\textregistered} Gorilla\textsuperscript{\textregistered} glass sample after traveling through an NA-0.4 microscope objective with long working distance.\cite{flamm2023photonic} After inscribing the modifications the good part is separated from the bad part by wet chemical etching with 30 wt-\% KOH \cite{kaiser2019selective} inside a heated bath. The resulting sample with the 45$\,^{\circ}$ chamfer is shown in Fig.~\ref{fig:FotoShowSample}, together with laser scanning microscope and scanning electron microscope images. The surface roughness was determined to $S_a \sim \unit[1]{\upmu m}$. For comparing the processed sample with our etch model, cf.~Sec.~\ref{sec:sims}, the LSM-measured cross section is averaged in $y$-direction for noise reduction purposes. The etch simulation covers $\approx 2000$ discrete time steps corresponding to an overall etch time of $30\,\mathrm{min}$, where the parameter $B$ was adjusted so, that $\approx 2\,\upmu\mathrm{m}$ are etched per minute. In the simulation, the selectivity of the modifications is assumed to be $18$ and the Super-Gaussian exponents of $N=4$ for type-III-regime modifications with diameters of $w_\mathrm{y}=2.5\,\upmu\mathrm{m}$ in transversal and $w_\mathrm{z}=13\,\upmu\mathrm{m}$ in propagation direction. In Fig.~\ref{fig:SimulationResults2D}, three different signals are plotted in five different time steps from $t = \unit[\left(45 \dots 1800\right)]{s}$. With the red curve, again, we show the measured edge height profile from our glass sample, cf.~Fig.~\ref{fig:FotoShowSample}. Highlighted in yellow, the modified glass region is denoted where we expect an etchable chain of voids, cf.~Fig.~\ref{fig:EtchingModel}. The light purple area shows the spatial progress of the etch solution penetrating the substrate. At time $t = \unit[1351]{s}$ the glass is completely etched as the solution has progressed to the substrate's center from both sides. Additionally, the benefit of the two assist etch lines becomes visible. The good part of workpiece is no longer interlocked in the substrate and can be detached upwards/downwards. This approach is mandatory if complex internal contours are to be extracted.\cite{kleiner2023laser}
\par
\begin{figure}
    \centering
    \includegraphics[width=0.5\textwidth]{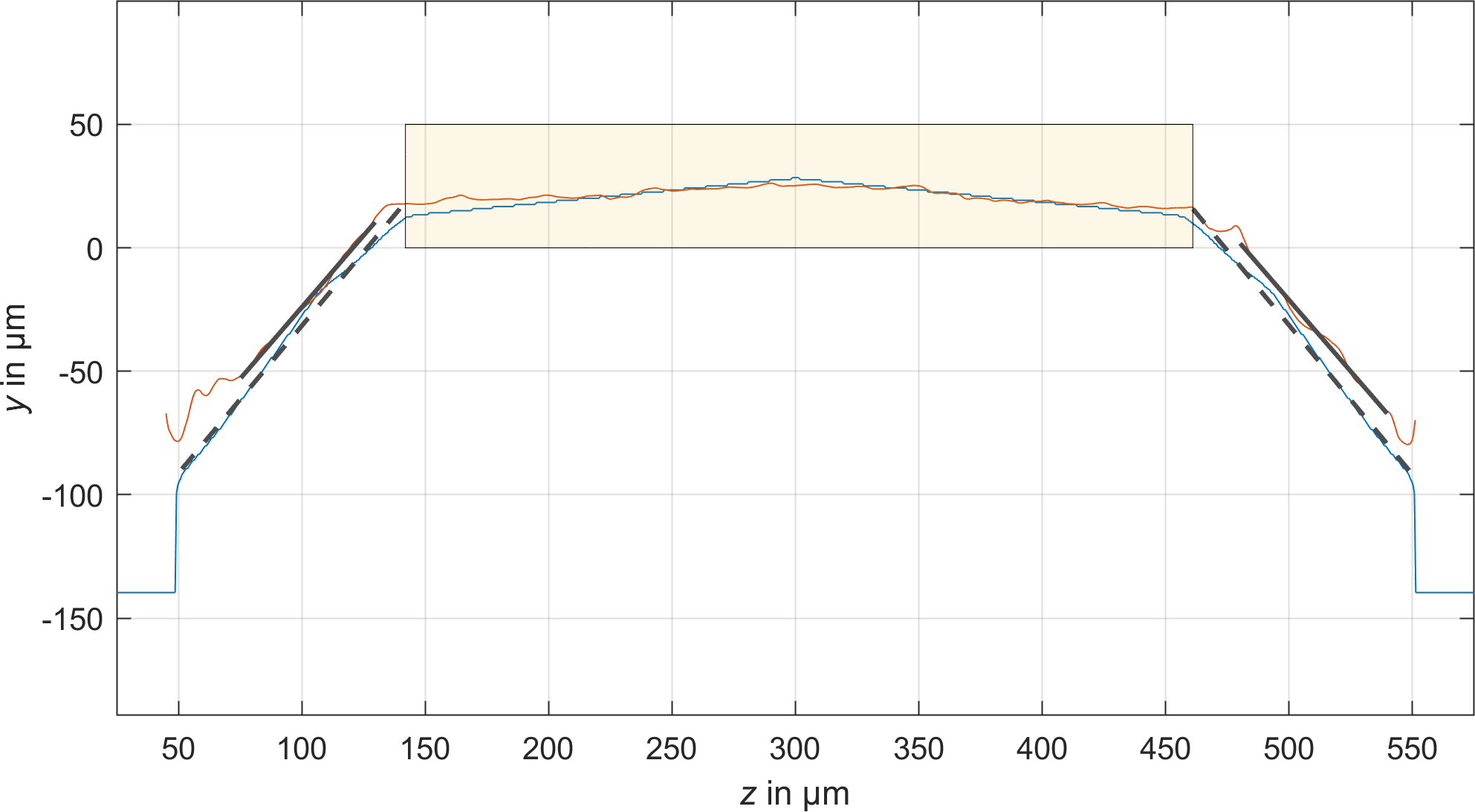}\vspace{1mm}
    \caption{Comparison of simulation and measurement of the chamfer as shown in Fig.~\ref{fig:SimulationResults2D}. The gray lines illustrate the regions of the linear fit in the experiment (solid) and simulation (dashed lines). Within the highlighted box, the standard deviation between simulation and experiment is about $2\,\upmu\mathrm{m}$ mainly caused by a slowly varying overall shape deviation.}
    \label{fig:ComparsionSimulationMeasurement}
\end{figure}

The overall and mid-frequent shape of the chamfer is very well reproduced in simulations and fits to the experimental results overlayed in Fig.~\ref{fig:SimulationResults2D} and shown in detail in Fig.~\ref{fig:ComparsionSimulationMeasurement}. Higher frequent deviations result from noise in the LSM measurement as well as deviations in the etching properties of individual type-III-regime modifications. 
The final glass thickness in the simulation is $503\,\upmu\mathrm{m}$ and fits well to the experimental values. For further characterizing the quality of the experiments, we evaluate the angles of the chamfer on the top and bottom side, which are $68.4\,^{\circ}$ and $66.3\,^{\circ}$ respectively and $68.6\,^{\circ}$ and $69.3\,^{\circ}$ in the experiment. Due to the symmetry of the simulation these values to be identical up to deviations in fitting a linear function in the region of the chamfer. Very near the edge of the sample, the light scanning microscope measurements show pronounced noise, which strongly influences the fitting. This is also visible in the central panel of Fig.~\ref{fig:FotoShowSample}. Therefore, the angle of the chamfer in the experiment is evaluated by fitting a linear function within a limited region depicted by the length of the solid gray line in Fig.~\ref{fig:ComparsionSimulationMeasurement}. Additionally, we evaluated the standard deviation between experiment and simulation within the central part of the sample (highlighted area in the figure) and find mainly slowly varying shape deviations to be the root cause for the $2\,\upmu\mathrm{m}$ difference. Within the tilted planes of the chamfer there is a parallel shift of the chamfer visible.

\section{Expansion to 3D}
\begin{figure}
    \centering
    \hfill
    \begin{subfigure}{3cm}
        \includegraphics[width=\textwidth]{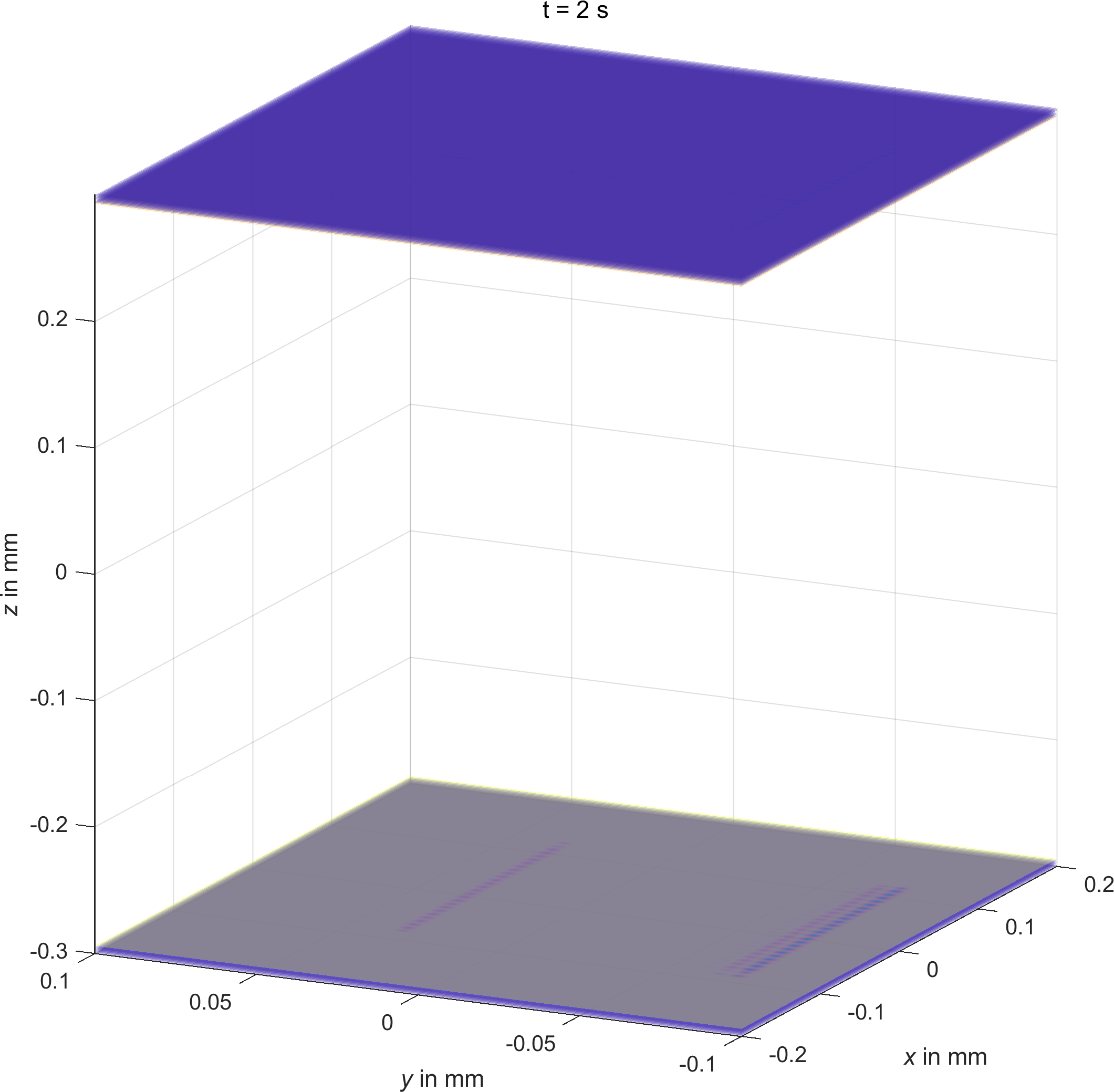}
    \end{subfigure}
    \hfill
    \begin{subfigure}{3cm}
        \includegraphics[width=\textwidth]{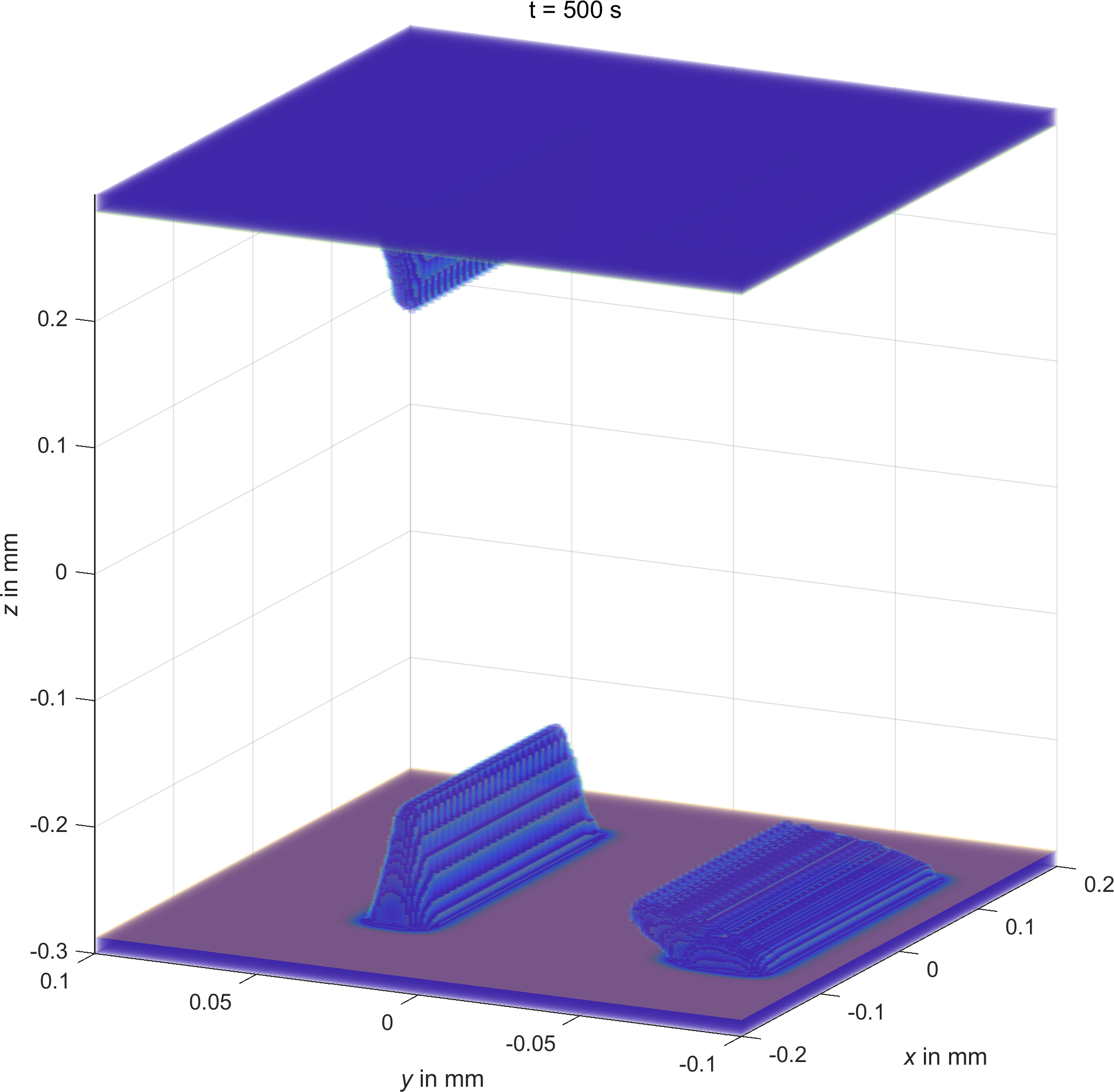}
    \end{subfigure}
    \hfill
    \begin{subfigure}{3cm}
        \includegraphics[width=\textwidth]{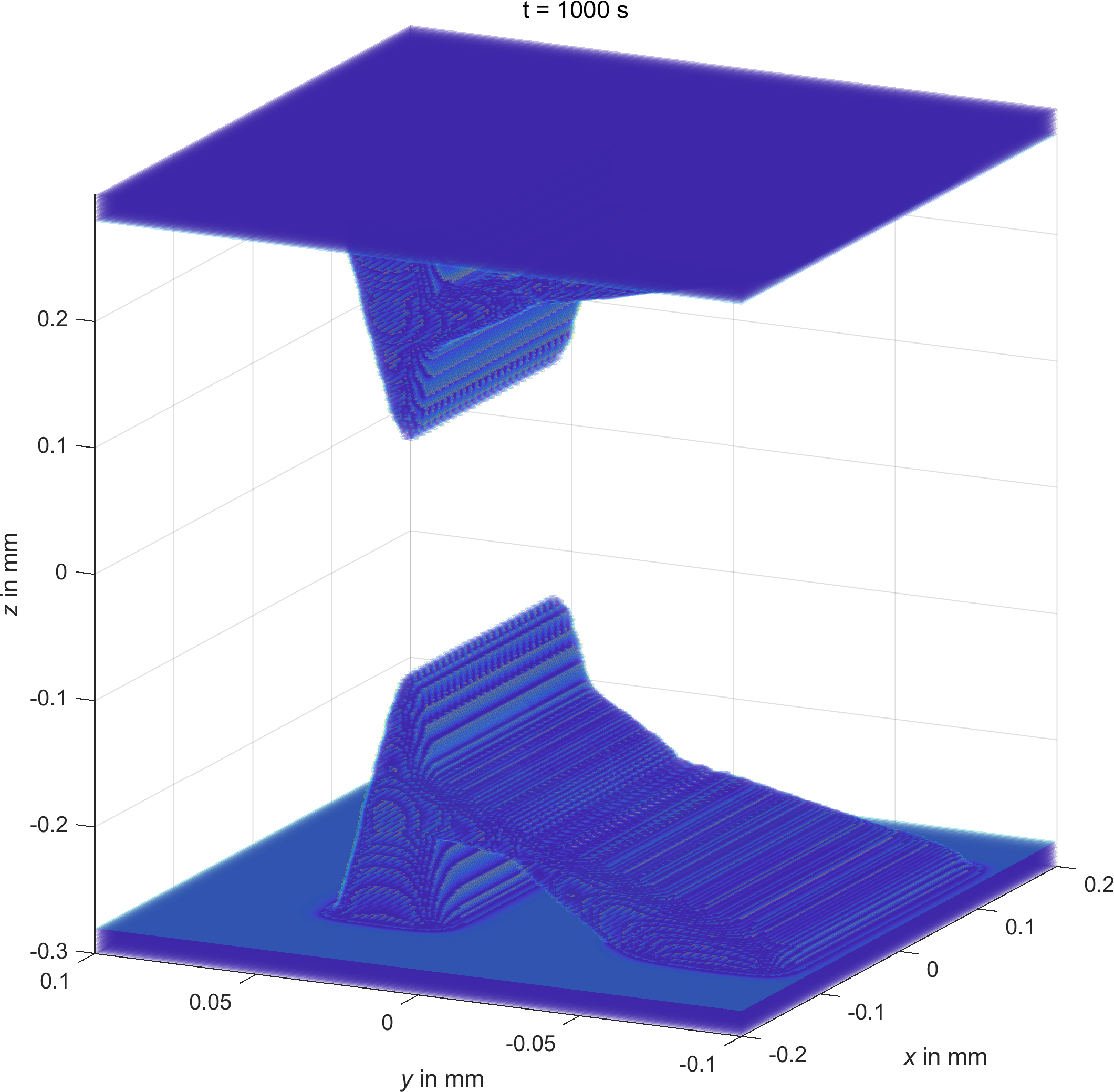}
    \end{subfigure}
    \hfill
    \begin{subfigure}{3cm}
        \includegraphics[width=\textwidth]{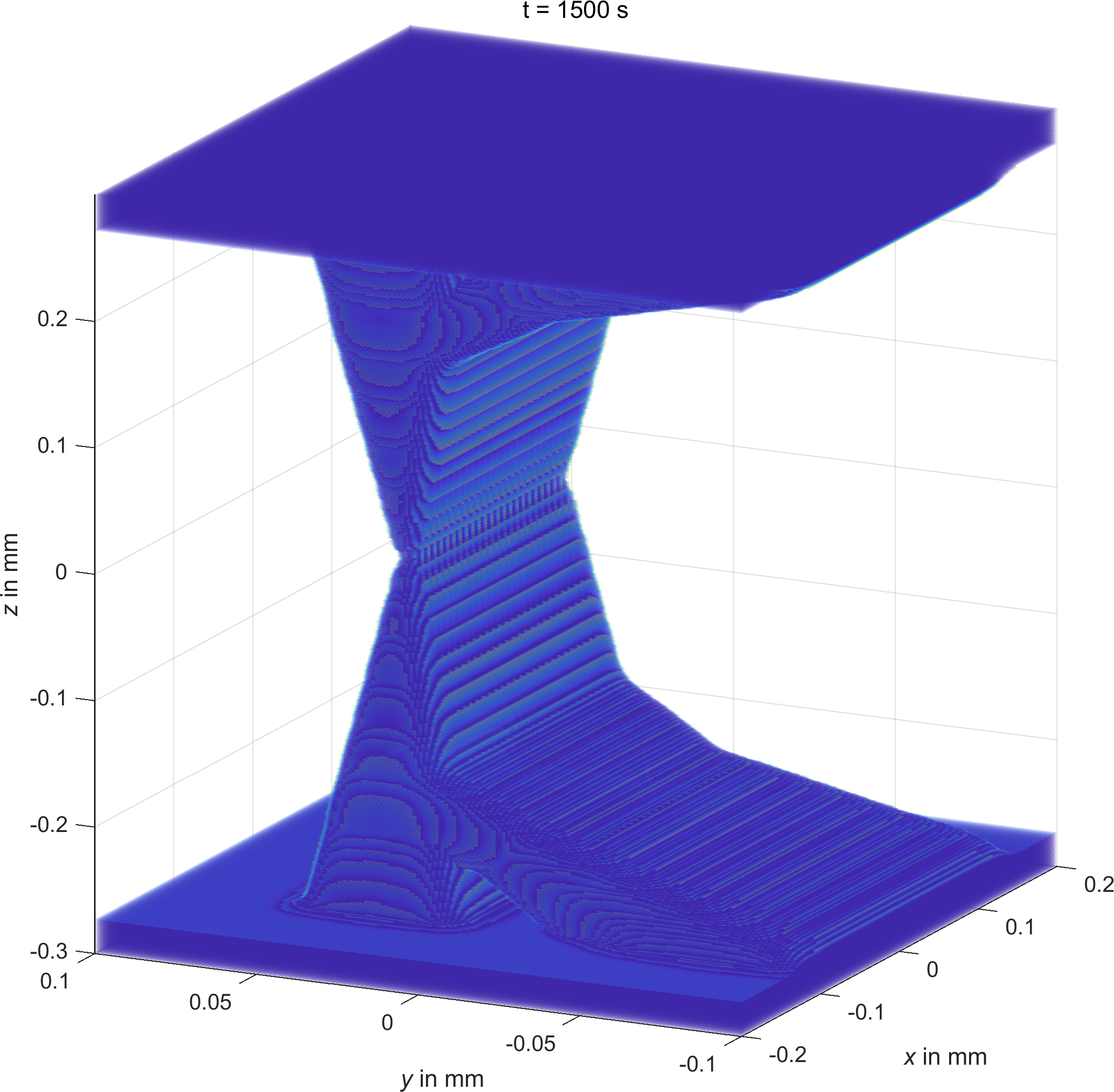}
    \end{subfigure}
    \hfill
    \begin{subfigure}{3cm}
        \includegraphics[width=\textwidth]{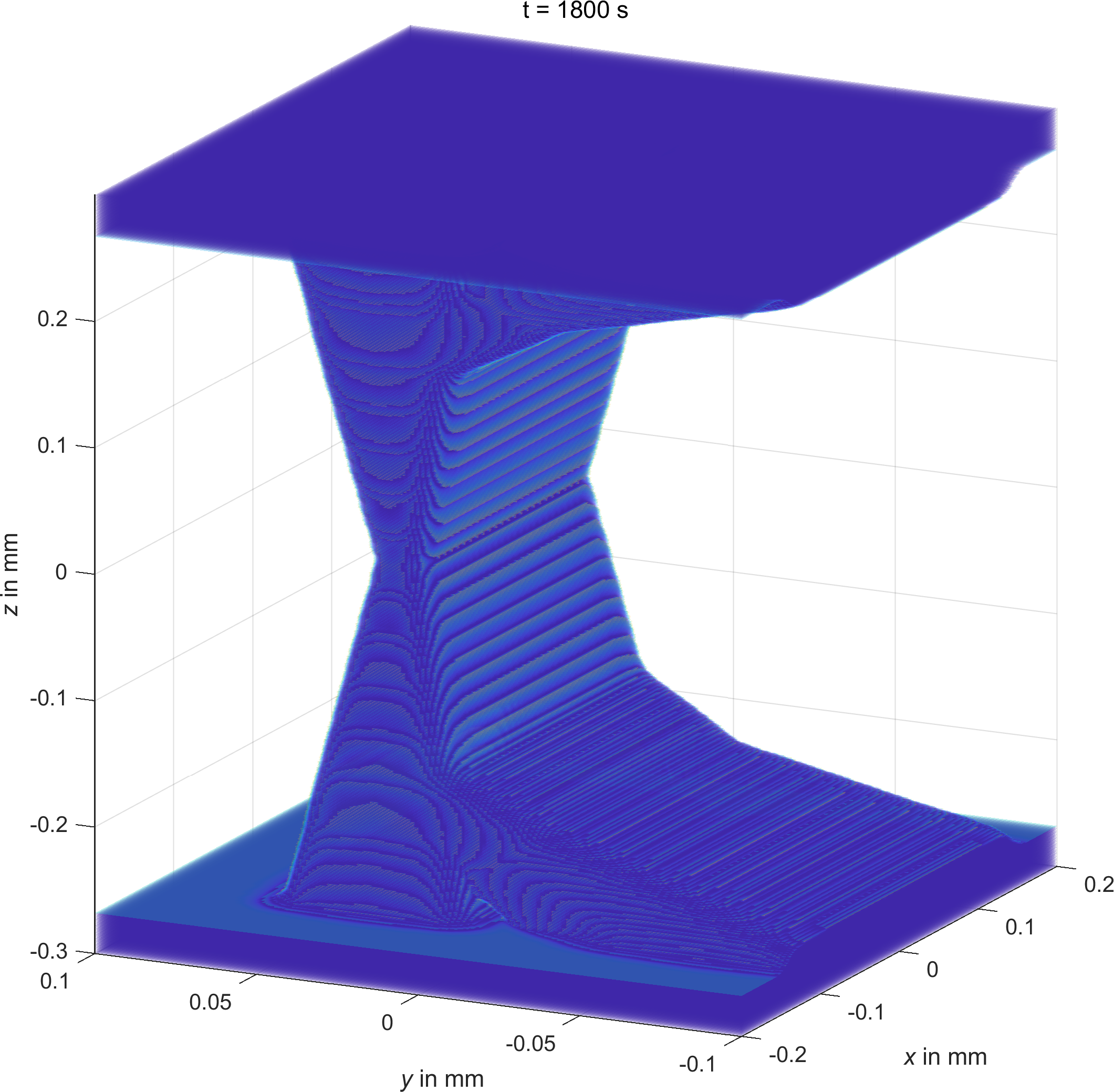}
    \end{subfigure}
    \hfill \vspace{1mm}
    \caption{Similar to Fig.~\ref{fig:SimulationResults2D} different temporal snapshots of a full 3D selective etch simulation are presented for the same chamfer geometry as in 2D. A feed in $x$-direction completes the 3 dimensional space.}
    \label{fig:SimulationResults3D}
\end{figure}

As indicated in Eq.~\eqref{eq:MeshEq3D}, the same ansatz as in 2D can be expanded to three spatial dimensions. Due to the cellular automaton the computational time for the 3D simulations is still small compared to a numerical simulation of a 3D reaction diffusion equation. This enables even more complex simulations including the feed direction of the sample as shown in Fig.~\ref{fig:SimulationResults3D}.

\begin{figure}
    \centering
    \includegraphics[width=1\textwidth]{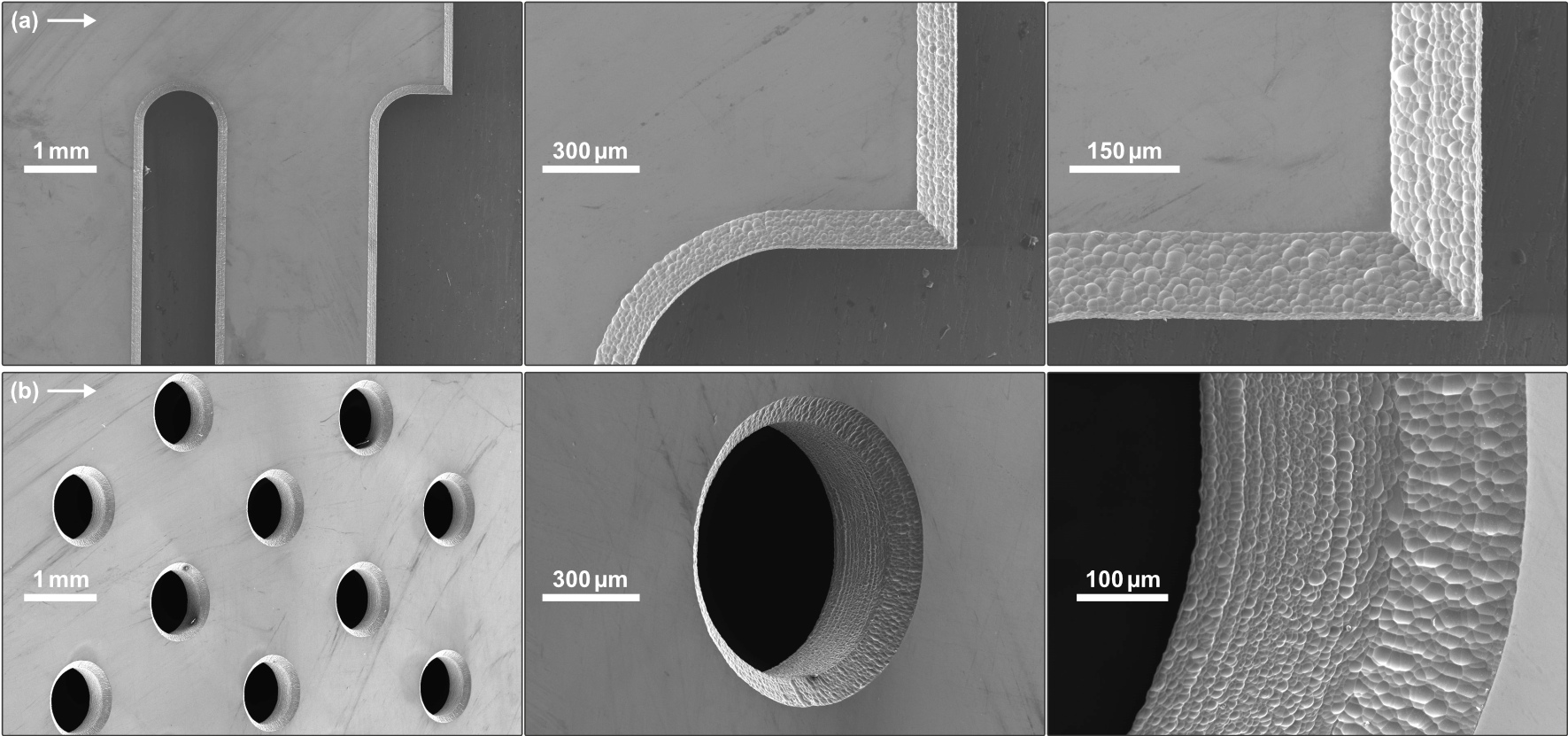}\vspace{1mm}
    \caption{Scanning electron microscope signals of a $\unit[550]{\upmu m}$-thick Corning\textsuperscript{\textregistered} Gorilla\textsuperscript{\textregistered} glass sample processed with a holographic 3D beam splitter and selective laser etching. The subfigures show complex outer (a) and inner (b) contours with different magnifications. A photograph of the total sample is provided in Fig.~\ref{fig:show}.}
    \label{fig:show}
\end{figure}

Such simulations allow to optimize focus shapes in cases where complex contours are to be etched and especially when inner contours need to be detached.\cite{kleiner2023laser}
Examples of this could be small holes or notches that are required in cover glasses in a wide variety of designs. In Fig.~\ref{fig:show}, a selection of possible contours is illustrated. Fabricated tailored-edge through-glass-vias exhibit a diameter of $\unit[1]{mm}$ (not the lower limit).

\section{Conclusion}
In summary, we demonstrated a conclusive tool chain starting with the optical process of inscribing modifications with ultrashort pulse lasers into glass samples followed by our numerical approach of evaluating the etching behavior including a forecast of the etched geometry. We find our numerical results to be in very good agreement with the experimental results. Based on the etching simulations it is possible to further increase the convergence between theoretical design of a chamfer and processed glass samples. In this way, we increase the quality of etched glass chamfers and pave the ground towards more complex chamfer geometries or higher requirements on the chamfer.

\bibliographystyle{spiebib}   
\bibliography{Lit}

\end{document}